\documentclass[sort&compress]{IEEEtran}
\usepackage{amsmath,amssymb}
\usepackage{multicol,multirow,diagbox}
\usepackage{graphicx,subfigure,float} 
\usepackage{lineno,cases}
\allowdisplaybreaks 
\usepackage{epstopdf} 
\usepackage{pgf,tikz}
\usetikzlibrary{arrows,shapes,automata,positioning}
\usetikzlibrary{fit}
\usepackage{pst-node}
\usepackage{color,xcolor}

\usepackage[pdfstartview=FitH,
CJKbookmarks=true,
bookmarksnumbered=true,
bookmarksopen=true,
colorlinks,
pdfborder=001,
linkcolor=blue,
anchorcolor=blue,
citecolor=blue
]{hyperref}

\renewcommand{\top}{{\mathrm{T}}}
\usepackage{bm} 
\newcommand{\lm}{\lambda}

\newcommand{\R}{\mathbb{R}}
\newcommand{\N}{\mathcal{N}}
\newcommand{\A}{\mathcal{A}}
\newcommand{\pb}{\begin{IEEEproof} }
\newcommand{\pe}{\end{IEEEproof}}

\newtheorem{lem}{Lemma}
\newtheorem{ass}{Assumption}
\newtheorem{thm}{Theorem}
\newtheorem{rem}{Remark}

\newtheorem{cor}{Corollary}
\begin{document}

\title{Event-triggered Design for Optimal Output Consensus of High-order Multi-agent Systems 
	\thanks{This work has been submitted to the IEEE for possible publication. Copyright may be transferred without notice, after which this version may no longer be accessible.}
}
\author{Yutao Tang, Huaihui Liu,  Ruonan Li, Kui Zhu
	\thanks{This work was supported by the National Natural Science Foundation of China under Grants 61973043.  Y. Tang, H. Liu, R. Li, and K. Zhu are with the School of Artificial Intelligence, Beijing University of Posts and Telecommunications, Beijing 100876, China. (E-mails: yttang@bupt.edu.cn,hui\_liu@bupt.edu.cn, nanruoliy@163.com, kuizhu\_19@bupt.edu.cn)}
}	
\date{ }
\maketitle

\begin{abstract}
	This paper studies the optimal output consensus problem for a group of heterogeneous linear multi-agent systems. Different from existing results, we aim at effective controllers for these high-order agents under both event-triggered control and event-triggered communication settings. We conduct an embedded design for the problem and constructively propose a multi-rate event-triggered controller with a set of applicable parameters. The proposed event-triggered rules are shown to be free of Zeno behaviors and can achieve the optimal output consensus goal for these high-order agents. A simulation example is given to verify the efficacy of our designs. 
\end{abstract}
	

\IEEEpeerreviewmaketitle

\section{Introduction}

A recent hot topic in multi-agent coordination is to develop effective rules for high-order agents  to achieve an optimal consensus specified by some distributed optimization problem \cite{yang2019survey}. This problem is a natural extension of existing distributed optimization results for single integrators to non-trivial dynamic agents and has been found in many practical scenarios especially when some engineering  multi-agent systems are involved \cite{li2014cooperative, zhang2017distributed}. Since the physical plants are often described by continuous-time models, many efforts have been made to develop continuous-time algorithms to achieve the optimal consensus goal for  different classes of high-order multi-agent systems, e.g., \cite{xie2017global, tang2019optimal, tang2021optimal, an2021distributed, wang2020distributed, liu2022new, li2022exponential}.

In the digital implementation of such continuous-time algorithms,  the controller update and information sharing are only allowed to happen at some discrete time instants instead of the whole time interval $[0,\,\infty)$. A typical treatment is to develop their time-scheduled periodic counterparts, i.e., the controller and shared information are updated periodically according to some (different) fixed periods.  Nevertheless, such algorithms may be inefficient from a computational perspective as the periods for control execution and information sharing are usually determined by some worst-case analysis. 

At the same time, event-triggered mechanisms have been extended to  multi-agent systems in different circumstances to alleviate the computation and communication burden \cite{heemels2012introduction, nowzari2019event}. Depending upon which action the agents take, we may divide event-triggered designs into two different categories as that in \cite{nowzari2019event}, i.e., event-triggered control and event-triggered communication.  In such event-triggered designs, the agents do not have to update its controller or share its own information with other agents periodically. These actions are taken only if they are necessary according to some extra real-time aperiodic mechanisms.  For example,  \cite{kia2015distributed} considered the optimal consensus/distributed optimization problem for single integrators under discrete (event-triggered) communication conditions.  Similar designs were delivered in \cite{liu2016event, deng2016distributed, kajiyama2018distributed, liu2019distributed, li2020distributed} to meet different optimization requirements. Note that these results were derived for single-integrator multi-agent systems. 
Some recent interesting attempts were made for second-order or linear agents in \cite{yi2018distributed, wang2018event, wang2019distributed, li2019distributed, yu2021new} . However, most existing event-triggered designs are only limited to the communication aspect assuming the actuator/controller of each agent can be continuously updated and instantly react to the received information. When the agents are physical and high-order ones, it is crucial for us to take the controller update issue into consideration to avoid unnecessary computation and energy consumption.   

Motivated by this observation, we consider the optimal consensus problem for heterogeneous linear multi-agent systems by event-triggered designs. Particularly, we will emphasize on both the control and communication sides to solve the problem and develop effective distributed controllers under both event-triggered control and event-triggered communication settings. In a preliminary version of this work \cite{tang2021ccc}, we discussed the event-triggered control issue but with continuous-time communication. By contrast, the controller update and information sharing here do not have to be implemented in a continuous or synchronously discrete fashion. Thus, the resulting controller is inherently asynchronous and of a multi-rate nature for the control and communication aspects. This will inevitably make the control design and performance analysis problem much more challenging than similar designs with either event-triggered control or event-triggered communication.

To solve the problem, we conduct an embedded control design as in \cite{tang2019optimal,tang2021optimal} and consider the construction of corresponding event-triggered optimal signal generator and event-triggered reference tracking controller separately.  We first introduce an optimal signal generator with arbitrarily fast exponential convergence rate. Then, we present an event-triggered version of the generator to reproduce the optimal solution under discrete communication. This together with an even-triggered tracking controller is shown to be Zeno-free and solve the formulated optimal consensus problem.  
 
The contribution of this paper is two-fold.  
\begin{itemize}
	\item We formulate and solve the optimal output consensus problem for heterogeneous multi-agent systems via event-triggered designs.  The derived results extended some existing optimal consensus works for integrator-type  multi-agent system to more general agent dynamics.  
	\item We provide novel multi-rate event-triggered designs for the optimal consensus problem. Compared with existing event-triggered communication results, we further take the controller update issue into consideration and can enlarge the potential application of such optimal consensus algorithms for engineering multi-agent systems.  
\end{itemize}

The rest of this paper is organized as follows:  We first state our problem in Section \ref{sec:form}. Then we present the main results of this paper in Section  \ref{sec:main}. Finally, an example is given in Section \ref{sec:simu} with some concluding remarks in Section \ref{sec:con}. 

{\sl Notations}: Let $\R^n$ be the $n$-dimensional Euclidean space and $\R^{n\times m}$ be the set of all $n\times m$ real matrices. ${\bf 1}_n$ (or ${\bf 0}_n$) denotes an $n$-dimensional all-one (or all-zero) column vector and ${\bm 1}_{n\times m}$ (or ${\bm 0}_{n\times m}$) all-one (or all-zero) matrix.  $\mbox{col}(a_1,\,{\dots},\,a_n) = {[a_1^\top,\,{\dots},\,a_n^\top]}^\top$ for column vectors $a_i\; (i=1,\,{\dots},\,n)$.  Let $M_1=\frac{1}{\sqrt{N}}{\bm 1}_N$ and $M_2$ be the matrices satisfying $M_2^\top M_1={\bm 0}_{N-1}$, $M_2^\top M_2=I_{N-1}$, and $M_2 M_2^\top=I_{N}-M_1 M_1^\top$.   For a vector $x$ (or matrix $A$) , $||x||$ ($||A||$) denotes its Euclidean (or spectral) norm.  

\section{Problem Statement}\label{sec:form}

Consider a high-order multi-agent system of the form:
\begin{align}\label{sys:agent}
    \begin{split}
        \dot{x}_i&=A_ix_i+B_iu_i\\
        y_i&=C_ix_i,  \quad i=1,\,\dots,\,N
    \end{split}
\end{align}
where $x_i\in \R^{n_i}$, $u_i\in \R$, and $y_i\in \R$ are the state, input, and output variables of agent $i$. System matrices $A_i,\, B_i,\, C_i$ are constant with proper dimensions. Moreover, we assume the triple $(C_i,\, A_i,\, B_i)$ is minimal without loss of generality. 

Suppose each agent $i\in \mathcal{N}$ has a twice continuously differentiable cost function $f_i \colon \mathbb{R} \to \mathbb{R}$ and let $f(s)=\sum\nolimits_{i=1}^Nf_i(s)$ be the global cost function. We aim at distributed rules to achieve an optimal output consensus for this multi-agent system on the minimizer of function $f$.

To this end, we assume the agents can share information with others through a topology described by a weighted graph $\mathcal{G}= \{\mathcal{N},\, \mathcal{E},\,\mathcal{A}\}$ with node set $\mathcal{N} \triangleq \{1,\,\dots,\, N\}$, edge set $\mathcal{E}\subset \N\times \N$, and weighted adjacency matrix $\A\in \R^{N\times N}$. The entry $a_{ij}$ at the $i$-th row and $j$-th column of $\mathcal{A}$ is strictly positive if there exists an directed edge from node $j$ to node $i$ and is $0$ otherwise. Here, a directed edge $(i,\,j)\in \mathcal{E}$ means agent $j$ can get access to the information of agent $i$. Let $\mathcal{N}_i=\{j\in \N\mid (i,\,i)\in \mathcal{E}\}$ be the set of all immediate neighbors of agent $i$.  More graph notations can be found in \cite{mesbahi2010graph}.  
 
Here are two standing assumptions to ensure the wellposedness of our problem.
\begin{ass}\label{ass:func}
	For any $i\in \N$ and $s\in \R$, it holds that $\underline{h}_i  \leq \nabla^2 f_i(s) \leq \bar h_i$ for some constants $\bar h_i\geq \underline{h}_i>0$.
\end{ass}
\begin{ass}\label{ass:graph}
	 $\mathcal{G}$ is strongly connected and weight-balanced.
\end{ass}

Under Assumption \ref{ass:func}, function $f$ has a unique minimal solution. 
Assumption \ref{ass:graph} ensures each agent's information can be reached by any other agent. Let $\mbox{Sym}(L)=\frac{L+L^\top}{2}$ with $L$ the Laplacian of digraph $\mathcal{G}$.  It is verified that $\mbox{Sym}(L)$ is positive semidefinite with all eigenvalues being real. Under Assumption \ref{ass:graph}, we can further order its eigenvalues as $0= \lambda_1 <\lambda_2 \leq \dots\leq \lambda_N$ . 
Then the optimal output consensus problem  for this multi-agent system \eqref{sys:agent} is to find a distributed controller $u_i$ compatible with graph $\mathcal{G}$ for agent $i$ such that $\lim_{t\to \infty} |y_i(t)-y_j(t)|\to 0$  and $\lim_{t\to \infty} y_i(t)= y^*$ for each $i\in \mathcal{N}$ with $y^*$ the minimal solution to function $f$.   

This problem has been partially studied under the name of distributed optimization/optimal coordination for different classes of high-order agents via continuous-time algorithms. 
Meanwhile, many authors proposed event-triggered implementation for the problem to avoid the possible inefficiency and save communication resources from the computational viewpoint. 
However, all the aforementioned works require the agent actuator to update in a continuous manner, which may be not necessary and might be prohibitive. When the agents are physical and high-order ones, it is favorable for us to determine some discrete time instants according to the real-time information and then adjust the controllers via some feedback mechanism to reduce the computational resources. Also, since both the actuators and sensors  can be updated at asynchronous discrete time instants in different varying rates, we are interested in distributed multi-rate rules with both event-triggered control and communication to solve the optimal output consensus  problem for  \eqref{sys:agent}.   

Before the main results, we make another assumption to ensure the solvability of our problem.  
\begin{ass}\label{ass:agent}
	For each $i\in \N$, system \eqref{sys:agent} has no transmission zeros at the origin. 
\end{ass}

This assumption specifies the class of high-order agents we are interested in. It is found to include the minimum-phase systems as special cases. Under this assumption, the so-called regulator equations $0=A_i X_i +B_i U_i,\, 1 =C_i X_i$ are ensured to have a unique solution pair $X_i$ and $U_i$ with proper dimensions \cite{huang2004nonlinear}.  Similar assumptions have been widely made in existing multi-agent coordination literature \cite{tang2019optimal, li2019distributed,  an2021distributed}.

\section{Main Result}\label{sec:main}

In this section, we solve the optimal output consensus problem for multi-agent system \eqref{sys:agent} via event-triggered designs. 

\begin{figure}
	\centering
	\includegraphics[width=0.42\textwidth ]{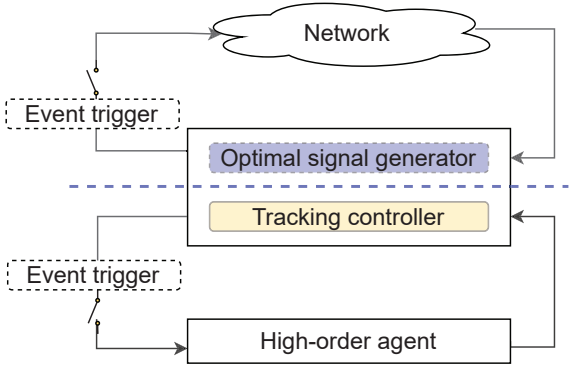}
	\caption{Embedded control scheme with event-triggered mechanisms.}\label{fig:diagram}
\end{figure}

We utilize the embedded design idea in \cite{tang2019optimal,tang2021optimal} and adapt it to the event-triggered setting. The basic structure is shown in Fig.~\ref{fig:diagram}. We divide the full problem into two layers. At the upper layer, we build an event-triggered optimal signal generator to reproduce the expected optimal consensus point. At the lower layer, we develop  an event-triggered tracking controller for each agent to converge to the generated estimate of the optimal solution. Bringing the two parts together, we will have the final distributed optimal consensus controller for the considered multi-agent system under both event-triggered control and event-triggered communication.

Let us start from the upper layer and introduce the following optimal signal generator: 
\begin{align}\label{sys:generator}
		\dot{z_i}&= -\alpha \nabla f_i(z_i)-\beta \sum_{j=1}^{N}a_{ij}(z_i- z_j)-\sum_{j=1}^{N} a_{ij}(v_i-v_j) \nonumber \\
		\dot{v_i} &=\alpha \beta \sum_{j=1}^{N}a_{ij}(z_i-z_j)
\end{align}
where $\alpha,\,\beta$ are positive constants to be specified later. 

Let $\underline{h}=\min_i\{\underline{h}_i\}$ and $\bar{h}=\max_i\{\bar h_i\}$.  The convergence of this generator relying on continuous communication has been established in \cite{tang2021optimal}.  Here we give a more elaborate property of this generator that its exponential convergence can be arbitrarily fast for some sufficiently large $\alpha$ and $\beta$. 

\begin{lem}\label{lem:generator}
	Suppose Assumptions \ref{ass:func} and \ref{ass:graph} hold. Let $\alpha\geq \max\{1,\, \frac{2\eta}{\min\{\underline{h},\,\lambda_2\}},\, \frac{6\bar h^2}{\underline{h}\lm_2}\}$ and $\beta\geq \max\{1,\, \frac{7\alpha^2\lambda_N^2}{\lambda_2^2} \}$ for any given $\eta>0$. Then, along the trajectory of the signal generator \eqref{sys:generator}, it holds that $|z_i(t)-y^*| e^{\eta t}< \infty$ for any $t\geq 0$ and $i\in \N$.
\end{lem}

Based on this lemma, we consider the implementation of  \eqref{sys:generator} with discrete communication.   For this purpose, we suppose each agent $i$ can broadcast its current information to its neighbors through the underlying communication network described by $\mathcal{G}$ at some discrete instants $\{\tilde t^i_k\}$ with $0=\tilde t^i_0<\tilde t^i_1< \dots$ for $i\in \mathcal{N}$ and $k\in \mathcal{Z}^+$. Then each agent updates the right-hand side of its generator instantly when receiving the most recent information from its neighbors. A traditional choice is to let $\tilde t_k^i=k\tau $ and consider the periodic communication to ensure the effectiveness of this generator. However, this setting suffers from the inefficiency issue as remarked before. Thus, we seek to event-triggering mechanism to determine the communication time instants to save resources. 


At the same time, we choose a matrix $K_{i1}$ such that $\hat A_i\triangleq A_i+B_iK_{i1}$ is Hurwitz and let $K_{i2}\triangleq U_i-K_{i1} X_i$. Then there exists a positive definite matrix $P_i$ such that $\hat A_i^\top P_i+P_i\hat A_i=-2I_{n_i}$. We denote by $\lm_{P_i}$ the maximal eigenvalue of $P_i$. 
According to the embedded control design procedure, we might substitute $y^*$ by its estimate $z_i(t)$ and utilize $\tilde u_i(t)=K_{i1} x_i(t)+K_{i2} z_i(t)$ for the lower layer to ensure an optimal consensus at $y^*$ for this multi-agent system \cite{tang2019optimal, tang2021optimal}.  

Motivated by this observation, we propose the final controller for agent $i$ with both event-triggered control and event-triggered communication as follows: 
\begin{align}\label{ctr:event-control-comm}
	u_i(t)&=K_{i1} x_i(t^i_k)+K_{i2} z_i(t^i_k)\nonumber\\
	\dot{z_i}&= -\alpha \nabla f_i(z_i)-\beta \sum_{j=1}^{N}a_{ij}(\tilde z_i- \tilde z_j)-\sum_{j=1}^{N} a_{ij}(\tilde v_i-\tilde v_j) \nonumber \\
	\dot{v_i} &=\alpha \beta \sum_{j=1}^{N}a_{ij}(\tilde z_i-\tilde z_j),\quad t\in [t^i_k,\,t^i_{k+1})  
\end{align}
with 
\begin{align*}
	\tilde{z}_i(t)=z_i(\tilde t_k^i),\quad \tilde{v}_i(t)=v_i(\tilde t_k^i),\quad t\in[\tilde t_k^i,\,\tilde t_{k+1}^{i})
\end{align*}
Here $\alpha$ and $\beta$ are chosen as above and the controller update time instants $\{t^i_{k}\}$ and communication time instants $\{\tilde t_{k}^i\}$ are determined by the following two rules: 
\begin{align} 
	t^i_{k+1}&=\inf\{t>t^i_k ~\big|~|\bar  u_i(t)|\geq c_0+c_1 e^{-\gamma t} \}\label{ctr:trigger-condition-ctr}\\
	\tilde t_{k+1}^{i}&=\inf \{ t>\tilde t_{k}^{i}~\big|~||\mbox{col}(\bar {z}_i(t),\bar{v}_i(t))|| \geq \tilde c_0+\tilde c_{1}e^{- \tilde \gamma t} \} \label{ctr:trigger-condition-comm}
\end{align}
with $\bar u_i=u_i-\tilde u_i$, $\bar z_i=z_i-\tilde z_i$, $\bar v_i=v_i-  \tilde{v}_i$, and  $c_0$, $c_1$, $\gamma$, $\tilde c_0$, $\tilde c_1$, $\tilde \gamma$ are nonnegative constants to be specified later.  This controller is distributed and essentially has multiple rates corresponding to  the event-triggered designs for both control and communication. 


Here is the main theorem of this paper. 
\begin{thm}\label{thm:main}
    Suppose Assumptions \ref{ass:func}--\ref{ass:agent} hold. Let $\alpha\geq \max\{1,\, \frac{2\eta}{\min\{\underline{h},\,\lambda_2\}},\, \frac{6\bar h^2}{\underline{h}\lm_2}\}$, $\beta\geq \max\{1,\, \frac{7\alpha^2\lambda_N^2}{\lambda_2^2} \}$ for some given constant $\eta>0$. Set $\tilde c_0\geq 0$, $\tilde c_1\geq 0$, $\tilde c_0+\tilde c_1>0$ with $0<\tilde \gamma<\min\{1,\,\frac{\eta }{2}\}$ for rule \eqref{ctr:trigger-condition-comm} and $c_0\geq \tilde c_0$, $c_1\geq 0$, $c_0+c_1>0$ with $0<\gamma<\min\{1,\,\frac{1}{2\lm_{P_i}},\, \tilde \gamma \}$ for rule \eqref{ctr:trigger-condition-ctr}. Then, along the trajectory of the composite system \eqref{sys:agent} and \eqref{ctr:event-control-comm} with triggering rules \eqref{ctr:trigger-condition-ctr}--\eqref{ctr:trigger-condition-comm},  the following assertions hold: 
    \begin{itemize}
    	\item[a)] There exists a constant $ \tau_{\min}>0$ such that   $t^i_{k+1}-t^i_k>{\tau}_{\min}$ and $\tilde t^i_{k+1}-\tilde t^i_k>{\tau}_{\min}$  hold for all $i\in \N$ and $k\in \mathbb{Z}^+$.
    	\item[b)] $y_i(t)$ will converge into an interval centered at $y^*$ with a radius proportional to $c_0+\tilde c_0\geq 0$ for any $i\in \mathcal{N}$. 
    	\item[c)] When $c_0=\tilde c_0=0$, $y_i$ will exponentially converge to the exact optimal solution $y^*$ as $t$ goes to $\infty$ for any $i\in \mathcal{N}$. 
    \end{itemize} 
\end{thm}

To prove this theorem, we first give a lemma on the effectiveness of the optimal signal generator with event-triggered communication. Its proof can be found in Appendix. 
  
\begin{lem}\label{lem:generator-comm}
	Suppose Assumptions \ref{ass:func}--\ref{ass:graph} hold. Consider the following event-triggered optimal signal generator
	\begin{align}\label{sys:event-generator}
		\dot{z_i}&= -\alpha \nabla f_i(z_i)-\beta \sum_{j=1}^{N}a_{ij}(\tilde  z_i- \tilde z_j)-\sum_{j=1}^{N} a_{ij}(\tilde v_i-\tilde v_j) \nonumber \\
		\dot{v_i} &=\alpha \beta \sum_{j=1}^{N}a_{ij}( \tilde z_i-\tilde z_j)
	\end{align}
	with $\{\tilde t_{k}^{i}\}$ determined by rule \eqref{ctr:trigger-condition-comm}. 
	Suppose the parameters are chosen as in Theorem \ref{thm:main}. 
	Then, there exists a constant ${\tilde \tau}_{\min}>0$ such that $\tilde t^i_{k+1}-\tilde t^i_k >\tilde \tau_{\min}$ holds for all $i\in \N$ and $k\in \mathbb{Z}^+$. Moreover, along the trajectory of  \eqref{sys:event-generator}, $z_i(t)$ exponentially converges into an interval centered at $y^*$  with a radius proportional to $\tilde c_0\geq 0$.  
\end{lem}

With this lemma, we are going to prove Theorem \ref{thm:main}. 

{\it Proof of Theorem \ref{thm:main}}: We first show the latter two items and then identify some ${\tau}_{\min}$ for item a) to complete the proof.

%
%
%

The composite system under our event-triggered control \eqref{ctr:event-control-comm} can be written as follows:
\begin{align*}
	\dot{x}_i&=\hat A_ix_i+B_iK_{i2} z_i+B_i\bar u_i \nonumber\\
	\dot{z_i}&= -\alpha \nabla f_i(z_i)-\beta \sum_{j=1}^{N}a_{ij}(\tilde  z_i- \tilde z_j)-\sum_{j=1}^{N} a_{ij}(\tilde v_i-\tilde v_j) \nonumber \\
	\dot{v_i}&=\alpha \beta \sum_{j=1}^{N}a_{ij}(\tilde z_i- \tilde z_j),\quad t\in [t^i_k,\,t^i_{k+1}),~~ k\in \mathbb{Z}^+ 
\end{align*}
We let $\bar x_i=x_i-X_i z_i$ and have 
\begin{align*}
	\dot{\bar x}_i&=\hat A_i \bar x_i+ B_i\bar u_i -X_i \dot{z}_i
\end{align*}
Set $V_i(\bar x_i)=\bar x_i^\top P_i \bar x_i$. Taking its time derivative along the trajectory of the above system, we have 
\begin{align*}
	\dot{V}_i&=2\bar x_i^\top P_i [\hat A_i \bar x_i+ B_i\bar u_i -X_i \dot{z}_i]\nonumber \\
	&\leq -2||\bar x_i||^2+2||\bar x_i||||P_iB_i|||\bar u_i||+2||\bar x_i||||P_iX_i||||\dot{z}_i| \nonumber \\
	& \leq -||\bar x_i||^2+ 2||P_iB_i||^2|\bar u_i|^2+ 2||P_iX_i||^2|\dot{z}_i|^2
\end{align*}
Then, we use the fact that $|\dot{z}_i|^2=|\dot{\tilde z}_i|^2$ and the Lipschitzness of the right-hand side of \eqref{sys:osg-reduced-event} and obtain 
\begin{align*}
	||\dot{z}_i||^2 &\leq 2\tilde h_M^2 ||\mbox{col}(\hat {z}(t),\hat{v}_2(t))||^2+2\tilde h_\Delta^2 ||\mbox{col}(\bar {z},\bar{v})||^2\\
	&\leq 4\tilde h_M^2 \alpha^3 W_0+2\tilde h_\Delta^2 N (\tilde c_0+\tilde c_1 e^{-\tilde \gamma t})^2
\end{align*}
Recalling the inequality \eqref{eq:lem-event-generator-W}, we further have 
\begin{align}\label{eq:thm-z}
	||\dot{z}_i||^2 & \leq 8\tilde h_M^2 \alpha^3 k_{W} e^{-2\tilde \gamma t} +2\tilde h_\Delta^2 N (\tilde c_0+\tilde c_1 e^{-\tilde \gamma t})^2 \nonumber\\
	&\leq (8\tilde h_M^2 \alpha^3 k_{W}+4\tilde h_\Delta^2 N\tilde c_1^2)  e^{-2\tilde \gamma t} + 4\tilde h_\Delta^2 N  \tilde c_0^2
\end{align}
This together with  the triggering mechanism implies that  
\begin{align}\label{eq1:thm-main-control}
	\dot{V}_i & \leq - ||\bar x_i||^2+ 4||P_iB_i||^2 c_0^2+4||P_iB_i||^2 c_1^2 e^{-2 \gamma t}\nonumber\\
	&+ 2 ||P_iX_i||^2 [(8\tilde h_M^2 \alpha^3 k_{W}+4\tilde h_\Delta^2 N\tilde c_1^2)  e^{-2\tilde \gamma t} + 4\tilde h_\Delta^2 N  \tilde c_0^2] \nonumber\\
	&\leq - \frac{1}{\lambda_{P_i}}V_i+k_\gamma e^{- 2\gamma t} + k_c (c_0+\tilde c_0)^2
\end{align}
for $k_\gamma \triangleq 16 \max_{i}\{ ||P_iX_i||^2 ( \tilde h_M^2 \alpha^3 k_{W}+ \tilde h_\Delta^2 N\tilde c_1^2)+ ||P_iB_i||^2 c_1^2 \}$ and $k_c=8\max_i\{  ||P_iB_i||^2+  ||P_iX_i||^2 \tilde h_\Delta^2 N \}$. Then $V_i(t)$ is uniformly ultimately bounded with an ultimate bound proportional to $(c_0+\tilde c_0)^2$ according to Lemma 9.8 in \cite{khalil2002nonlinear}. Recalling the definition of $V_i$ and Lemma \ref{lem:generator-comm}, one can directly conclude that $y_i-y^*=C_i\bar x_i+z_i-y^*$ will converge to an interval centered at $0$ with a radius proportional to $c_0+\tilde c_0$. This is exactly the statement in item b). 

Particularly, when $c_0= \tilde c_0=0$, we have 
\begin{align}\label{eq2:thm-main-control}
	V_i(t)&\leq V_i(0)e^{-\frac{t}{\lambda_{P_i}}}+ \int_{0}^t e^{-\frac{t-\tau}{\lambda_{P_i}}} k_\gamma e^{-2\gamma \tau } {\rm d}\,\tau \nonumber\\
	&\leq V_i(0)e^{-\frac{t}{\lambda_{P_i}}}+ k_\gamma  e^{-\frac{t}{\lambda_{P_i}}} \int_{0}^t e^{(\frac{1}{\lambda_{P_i}}-2\gamma)\tau } {\rm d}\,\tau \nonumber\\
	&\leq k_{V_i} [e^{-\frac{t}{\lambda_{P_i}}}+e^{-2\gamma t}]
\end{align}
with $k_{V_i} \triangleq V_i(0)+\frac{2\gamma\lambda_{P_i} k_\gamma }{1-2\lambda_{P_i} \gamma}$. This with Lemma \ref{lem:generator-comm} implies  the exponential convergence of $y_i$ towards $y^*$ as $t$ tends to $\infty$. 

Next, we will prove  item a).  We first determine such a constant for the controller part. For this purpose, we check the evolution of $|\bar u_i(t)|$.  Suppose agent $i$ is triggered to update its controller at $t^*\geq 0$ .  Before the next event, the derivative of $\bar u_i(t)$ satisfies that
\begin{align}\label{eq3:thm-main-control}
	\begin{split}
		\dot{\bar u}_i &=-\dot{\tilde u}_i =-(K_{i1}\dot{\bar x}_i+ U_i\dot{z}_i)\\
		&=-K_{i1}(\hat A_i\bar x_i+B_i\bar u_i)-K_{i2} \dot{z}_i
	\end{split}
\end{align}
From \eqref{eq:thm-z}, there must be some  $k_z>0$ such that $|\dot{z}_i|\leq k_z(\tilde c_0+\tilde c_1 e^{-\tilde \gamma t})$. We combine it with the fact that $|\bar u_i(t)|\leq c_0+c_1e^{-\gamma t}$ before the next triggering time and have   
\begin{align*}
	|\dot{\bar u}_i| & \leq ||K_{i1}\hat A_i \bar x_i||+||K_{i1}B_i||c_0+k_z ||K_{i2}|| \tilde c_0\\
	& + ||K_{i1}B_i|| c_1e^{-\gamma t} + k_z||K_{i2}||\tilde c_1e^{-\tilde \gamma t}\\
	&\leq ||K_{i1}\hat A_i \bar x_i||+ (||K_{i1}B_i|| c_1+k_z||K_{i2}||\tilde c_1)e^{-\gamma t}\\
	&+(||K_{i1}B_i||+k_z ||K_{i2}||)(c_0+\tilde c_0) 
\end{align*}
We consider two different cases to determine a lower bound for the inter-event intervals depending upon whether the constant $c_0$ is $0$ or not.

When $c_0\neq 0$. From the ultimate boundedness of $\bar x_i(t)$, there exists some $k_u>0$ such that $|\dot{\bar u}_i| \leq k_u$ holds for any $i\in \N$.  Hence, $|\bar u_i(t)|\leq \int_{t^*}^t  |\dot{\bar u}_i(t)|{\rm d}\,t \leq k_u (t-t^*)$. Note that the next event will not be triggered before $|\bar u_i(t)|$ reaches $c_0$.  Thus the inter-event interval must be lower bounded by $\tau_0=c_0/k_u$. 

When $c_0=0$, we have $\tilde c_0=0$. Recalling the inequality \eqref{eq2:thm-main-control}, we conclude that
\begin{align*}
	|\dot{\bar u}_i(t)|\leq \hat c_1 (e^{-\frac{t}{2\lambda_{P_i}}}+ e^{- \gamma t}) <2\hat c_1 e^{-\gamma t} 
\end{align*}
for some constant $\hat c_1>0$. Since $\bar u_i(t^*)=0$, it follows then 
\begin{align*}
	|\bar u_i(t)|\leq  2\hat c_1 e^{-\gamma t^*}(t-t^*)
\end{align*}
for $t>t^*$. By similar arguments as in the proof of Lemma \ref{lem:generator-comm}, we consider the equation $2\hat c_1s-c_1e^{-\gamma s}=0$ and denote its solution by $\tau_1>0$. For any $t^*\leq t<t^*+\tau_1$, it holds that $|\bar u_i(t)| <2\hat c_1 e^{-\gamma t^*}\tau_1=c_1e^{-\gamma (t^*+\tau_1)}<c_1e^{-\gamma t}$. This means the next event for rule \eqref{ctr:trigger-condition-ctr} is not triggered before $t^*+\tau_1$, or the inter-event interval is lower bounded by $\tau_1$.

Overall,  set $\tau_{\min}=\min \{\tau_0,\,\tau_1,\tilde \tau_{\min}\}$ for both cases with $\tilde \tau_{\min}$ determined in Lemma \ref{lem:generator-comm}. This confirms the statement in item a) and thus completes the proof. {\hfill\IEEEQEDhere }

\begin{rem}\label{rem:zeno}
	Theorem \ref{thm:main} states the effectiveness of the multi-rate event-triggered controller \eqref{ctr:event-control-comm} in solving the formulated optimal consensus problem. Moreover, the resulting closed-loop system  under this controller is free of Zeno behaviors.  
\end{rem}
\begin{rem}\label{rem:event}
	Different from most existing continuous-time (optimal) consensus designs relying on continuous or discrete communication \cite{nowzari2019event,xie2017global, tang2019optimal, deng2016distributed,  li2020distributed, wang2018event, li2019distributed, an2021distributed} , we considers both event-triggered control and event-triggered communication issues. Thus, the developed event-triggered controller removes the requirement of continuous controller update and continuous information sharing, which will definitely save many computational resources. 
\end{rem}

The preceding controller \eqref{ctr:event-control-comm} contains two parts, i.e., event-triggered   tracking controller and event-triggered optimal signal generator. If either continuous controller update or continuous communication is incorporated, we can get several special kinds of this distributed controller to solve the formulated optimal consensus problem. As the optimal consensus designs have been extensively studied with a continuous controller-updating requirement, we here present an invariant of \eqref{ctr:event-control-comm} to solve the problem without this requirement as follows:  
\begin{align}\label{ctr:event-control}
	u_i(t)&=K_{i1} x_i(t^i_k)+K_{i2} z_i(t^i_k)\nonumber\\
	\dot{z_i}&= -\alpha \nabla f_i(z_i)-\beta \sum_{j=1}^{N}a_{ij}(z_i-z_j)-\sum_{j=1}^{N} a_{ij}(v_i- v_j) \nonumber \\
	\dot{v_i} &=\alpha \beta \sum_{j=1}^{N}a_{ij}(z_i- z_j),\quad t\in [t^i_k,\,t^i_{k+1})  
\end{align}
with the controller update time $\{t^i_{k}\}$  determined by  \eqref{ctr:trigger-condition-ctr}.  

We summarize the effectiveness of this controller as follows and omit the proof to save space.  
\begin{cor} \label{cor:main-control}
	Suppose Assumptions \ref{ass:func}--\ref{ass:agent} hold. Let $\alpha\geq \max\{1,\, \frac{2\eta}{\min\{\underline{h},\,\lambda_2\}},\, \frac{6\bar h^2}{\underline{h}\lm_2}\}$, $\beta\geq \max\{1,\, \frac{7\alpha^2\lambda_N^2}{\lambda_2^2} \}$, and $0<\gamma<\min\{1,\,\frac{1}{2\lm_{P_i}},\, \eta \}$. Then, for any $c_0\geq 0$, $c_1\geq 0$, $c_0+c_1>0$, along the trajectory of the composite system \eqref{sys:agent} and \eqref{ctr:event-control} with \eqref{ctr:trigger-condition-ctr}, the following assertions hold: 
	\begin{itemize}
		\item[a)] There exists a constant $\bar \tau_{\min}>0$ such that $t^i_{k+1}-t^i_k>\bar \tau_{\min}$ holds for all $i\in \N$ and $k\in \mathbb{Z}^+$.
		\item[b)] $y_i(t)$ converges into an interval centered at $y^*$ with radius proportional to $c_0\geq 0$ for any $i\in \mathcal{N}$. 
		\item[c)] When $c_0=0$, $y_i$ will exponentially converge to the exact optimal solution $y^*$ as $t$ goes to $\infty$ for any $i\in \mathcal{N}$. 
	\end{itemize} 
\end{cor}


\section{Simulation} \label{sec:simu}
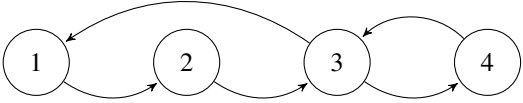
\begin{figure}
	\centering
	\normalsize
	\begin{tikzpicture}[shorten >=1pt, node distance=2 cm, >=stealth',
		every state/.style ={circle, minimum width=0.5 cm}, auto]
		\node[align=center,state](node1) {1};
		\node[align=center,state](node2)[right of=node1]{2};
		\node[align=center,state](node3)[right of=node2]{3};
		\node[align=center,state](node4)[right of=node3]{4};
		\path[->]   (node1) edge [bend right=35] (node2)
		(node3) edge [bend right=35] (node1)
		(node2) edge [bend right=35]  (node3)
		(node3) edge [bend right=35]  (node4)
		(node4) edge [bend right=45]  (node3)
		;
	\end{tikzpicture}
	\caption{Communication digraph $\mathcal{G}$ in our example.} \label{fig:graph}
\end{figure}

In this section, we consider a four-agent network to illustrate the effectiveness of our event-triggered controllers.  

\begin{figure} 
	\centering
	\includegraphics[width=0.42\textwidth]{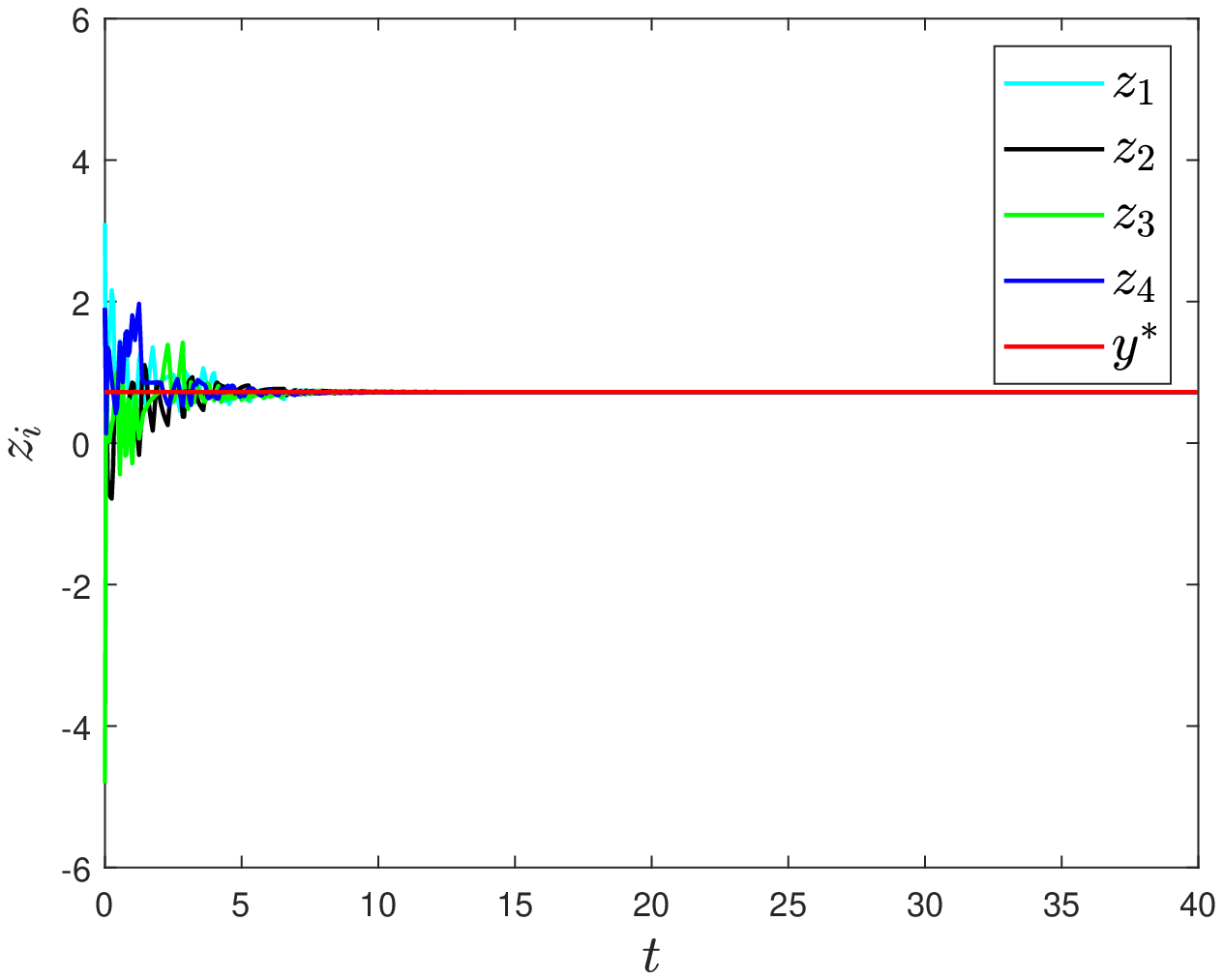}   
	\caption{Profile of  $z_i$ under the controller \eqref{ctr:event-control-comm}.} 
	\label{fig:simu-output0} 
\end{figure}

\begin{figure} 
	\centering
	\includegraphics[width=0.42\textwidth]{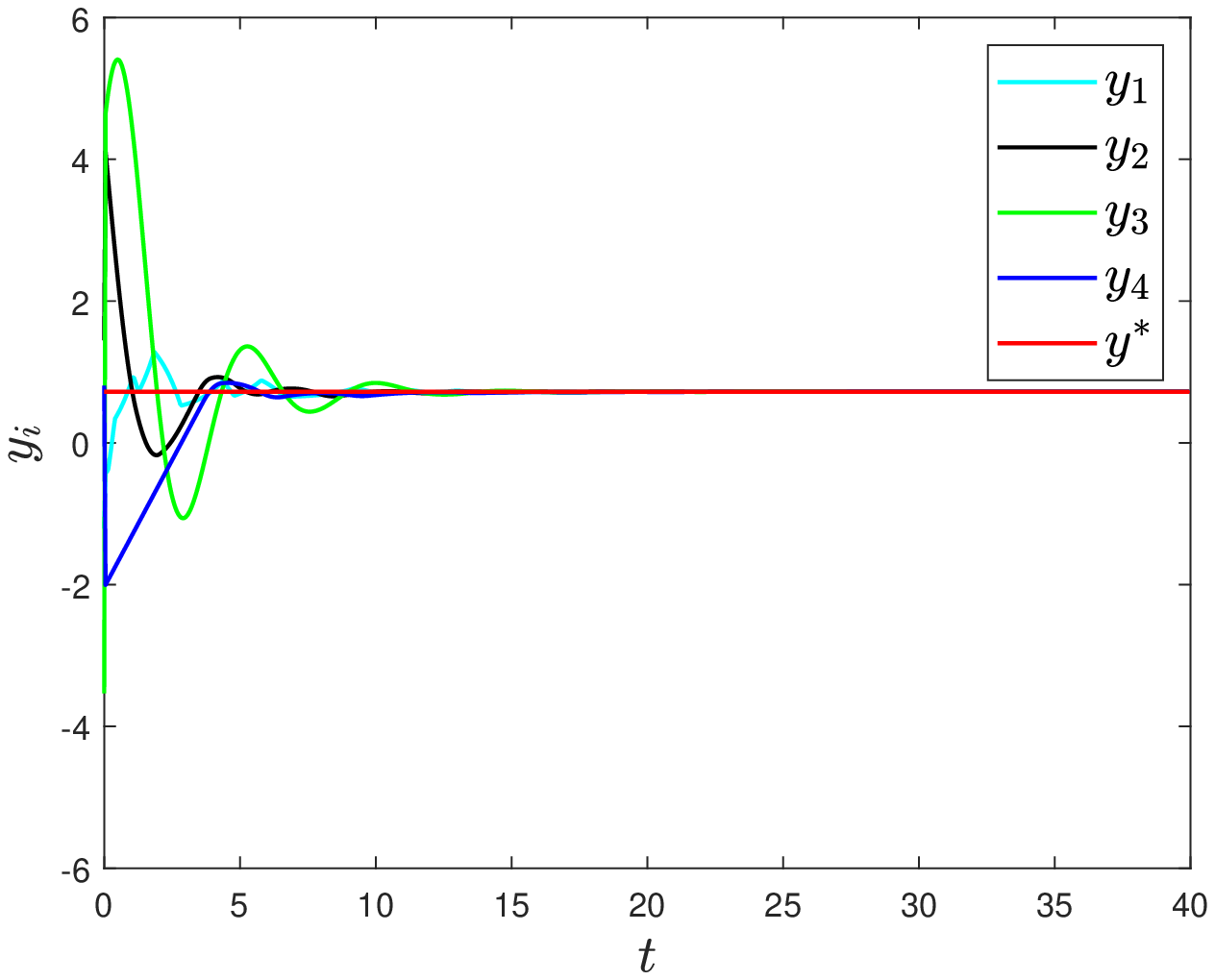}   
	\caption{Profile of agent outputs $y_i$ under the   controller \eqref{ctr:event-control-comm}.} 
	\label{fig:simu-output} 
\end{figure}

Suppose the system matrices are 
\begin{align*}
&	A_1=1,\; B_1=1,\; C_1=1\\
&	A_2=\begin{bmatrix}
		0&1\\-1&0
	\end{bmatrix},\; B_2=\begin{bmatrix}
	0\\ 1
\end{bmatrix},\; C_2= \begin{bmatrix}
1\\ 0
\end{bmatrix}^\top \\
&A_3=\begin{bmatrix}
		0&1&0\\
		-1&0&1\\
		2&0&1\\
	\end{bmatrix},\; B_3=\begin{bmatrix}
	0\\ 1\\1
\end{bmatrix},\; C_3=\begin{bmatrix}
0\\1\\0
\end{bmatrix}^\top \\
&	A_4=\begin{bmatrix}
		0&1\\0&0
	\end{bmatrix},\; B_4=\begin{bmatrix}
	0\\1
\end{bmatrix},\; C_4=\begin{bmatrix}
1\\0
\end{bmatrix}^\top 
\end{align*}
These agents are interconnected by a digraph depicted as in Fig.~\ref{fig:graph} with unity weights.  The cost functions are chosen as $f_1(s)=\frac{1}{2}(s-2)^2$, $f_2(s) = s^2 \ln(1 + s^2) + (s+1)^2$, $f_3(s)=\ln(e^{-0.1s}+e^{0.3s})+s^2$, and $f_4(s)=\frac{s^2}{25\sqrt{s^2 + 1}}+(s-3)^2$. Assumptions \ref{ass:func}--\ref{ass:agent} can be practically verified.

\begin{figure} 
	\centering
	\includegraphics[width=0.42\textwidth]{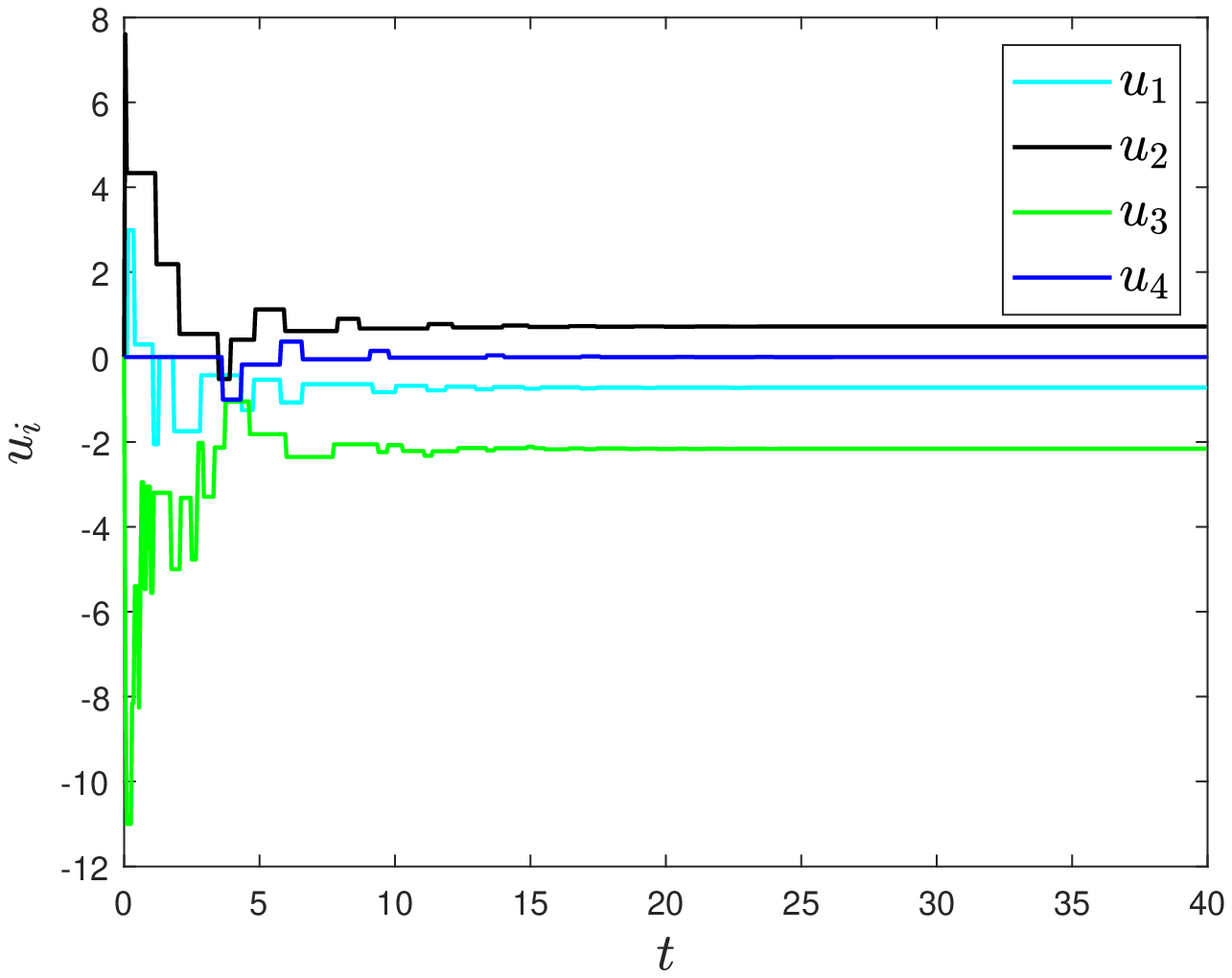}   
	\caption{Profile of agent inputs under the controller \eqref{ctr:event-control-comm}.} 
	\label{fig:simu-input} 
\end{figure}

\begin{figure} 
	\centering
	\includegraphics[width=0.42\textwidth]{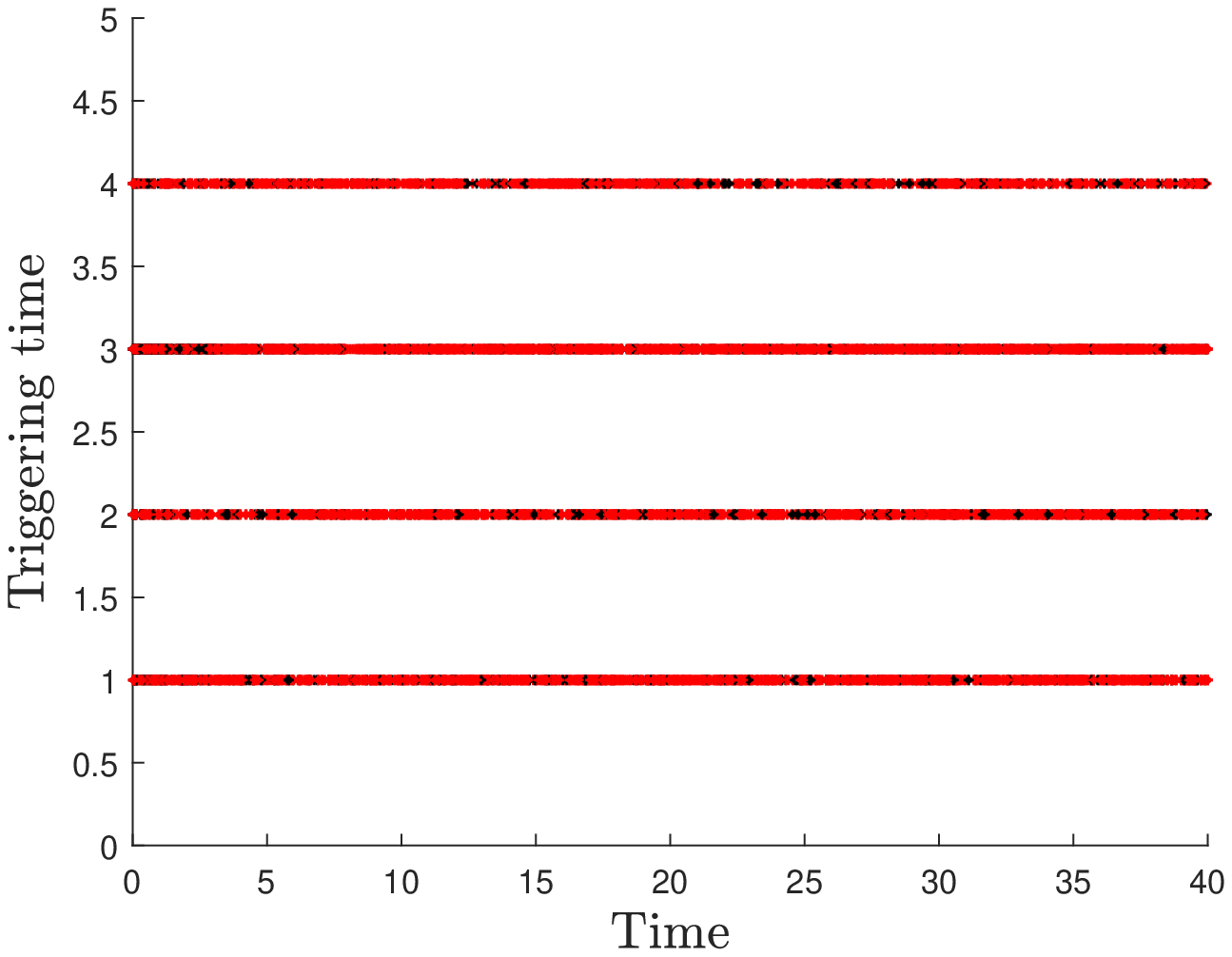} 
	\caption{Triggering time for each agent under the controller \eqref{ctr:event-control-comm}.} 
	\label{fig:simu-time} 
\end{figure}

In simulation, the control gain matrices are set as $K_{11}=-2.4142$, $K_{12}=1.4142$, $K_{21}=[ -0.4142~-1.3522]$, $K_{22}=1.4142$, $K_{31}=[-2.7331~-2.3372~-3.5835]$, $K_{32}=3.3166$, $K_{41}=[-1.0000~-1.7321]$, and $K_{42}=1.0000$. 
%
%
We let $\alpha=1$ and $\beta=10$ for the optimal signal generator and choose $c_0=\tilde c_0=0$, $c_1=\tilde c_1=5$, $\gamma=0.5$, and $\tilde \gamma=0.1$ for the event-triggered rules \eqref{ctr:trigger-condition-ctr}--\eqref{ctr:trigger-condition-comm}. The initial conditions are randomly chosen in $[-5,\,5]$. We list the profiles of $z_i$, $y_i(t)$, and $u_i(t)$ in Figs.~\ref{fig:simu-output0}--\ref{fig:simu-input}. It can be found that the outputs of all agents reach the expected consensus on the global optimal solution quickly while the control efforts are bounded and converge to some steady-states. We also show the triggering time of instants in Fig.~\ref{fig:simu-time} where the red markers are corresponding to even-triggered communication and the black ones corresponding to event-triggered control. These observations confirm the effectiveness of our event-based designs to solve the optimal output consensus problem for heterogeneous multi-agent systems.

%

\section{Conclusion}\label{sec:con}
We have considered the optimal consensus problem for a group of heterogeneous linear multi-agent systems via event-based designs. To solve this problem, we have presented an embedded design and provided a multi-rate event-triggered controller for this problem. Under some standing assumptions, we have shown that the developed controller is Zeno-free and can drive the agent outputs to reach an approximate or exact optimal consensus on the expected optimal point specified by the given optimization problem under different parameters.  

\appendix 

\subsection{Proof of Lemma \ref{lem:generator}}

We first put \eqref{sys:generator} into a compact form as that in \cite{tang2021optimal}: 
\begin{align*}
	\dot{z}&=-\alpha \nabla \hat f(z)- \beta Lz-Lv\\
	\dot{v}&=\alpha \beta L z
\end{align*}
where $\hat f(z)\triangleq \sum_{i=1}^N f_i(z_i)$ with $z=\mbox{col}(z_1,\,\dots,\,z_N)$ and $v=\mbox{col}(v_1,\,\dots,\,v_N)$. It can be verified that function $\hat f$ is $\underline{h}$-strongly convex while its gradient $ \nabla\hat f(z)$ is $\bar h$-Lipschitz. Let $\mbox{col}(z^*,\,v^*)$ be an equilibrium of this system. We can practically verify that $z^*={\bf 1}_N y^*$ under the lemma conditions. 

Perform the coordinate transformation: $\hat z_1=M_1^\top (z-z^*)$, $\hat z_2=M_2^\top (z-z^*)$, $\hat v_1=M_1^\top (v-v^*)$, and $\hat v_2=M_2^\top[( v+\alpha z)-( v^*+\alpha z^*)]$. It follows that $\dot{\hat v}_1= 0$ and  
\begin{align}\label{sys:osg-reduced}
	\begin{split}
		\dot{\hat z}_1&=-\alpha M_1^\top {\bm \Pi}\\
		\dot{\hat z}_2&=-\alpha M_2^\top {\bm \Pi}-\beta M_L \hat z_2 + \alpha M_L \hat z_2-M_L\hat v_2\\
		\dot{\hat v}_2&=-\alpha M_L {\hat v}_2+\alpha^2 M_L \hat z_2-\alpha^2 M_2^\top {\bm \Pi} 
	\end{split}
\end{align}
where ${\bm \Pi}\triangleq \nabla \hat f(z)-\nabla \hat f(z^\star)$ and $M_L=M_2^\top LM_2$.  Let $\hat z=\mbox{col}(\hat z_1,\,\hat z_2)$, and $W_0(\hat z,\,\hat v_2)=\frac{1}{2}\hat z^\top \hat z +\frac{1}{\alpha^3}\hat v_2^\top \hat v_2$ with the chosen constant $\alpha$ in the lemma. 
Meanwhile, following a similar procedure as that in \cite{tang2021optimal}, we take the time derivative of $W_0$ along the trajectory of \eqref{sys:osg-reduced} and have
\begin{align*}
	\dot{W}_0&\leq -\alpha\underline{h}||\hat z||^2- \beta\lambda_2||\hat z_2||^2+ \alpha  \lambda_N  ||\hat z_2||^2+ \lambda_N  ||\hat z_2||||\hat v_2||\\
	&-\frac{2\lambda_2}{\alpha^2}||\hat v_2||^2+ \frac{2\lambda_N}{\alpha}  ||\hat z_2|| ||\hat v_2||+\frac{2\bar h}{\alpha}||\hat v_2||||\hat z||\\
	&\leq -(\alpha\underline{h} -\frac{3\bar h^2}{\lambda_2} )||\hat z||^2-\frac{\lambda_2}{\alpha^2}||\hat v_2||^2\\
	& - (\beta\lambda_2-\alpha\lambda_N-\frac{3\alpha^2\lambda_N^2}{\lambda_2}-\frac{3\lambda_N^2}{\lambda_2})||\hat z_2||^2\\
	&\leq -2\eta W_0
\end{align*}
where we have used Young's inequality to handle the cross terms. Then, we solve this differential inequality and have $W_0(\hat z(t),\,\hat v_2(t))\leq W_0(\hat z(0),\,\hat v_2(0))e^{-2\eta t}$. This combined with the fact that $||z-z^*||=||\hat z|| $ implies that $$|z_i-y^*|e^{\eta t}\leq ||z-z^*||e^{\eta t}\leq  \sqrt{2W_0(\hat z(0),\,\hat v_2(0))}<\infty$$ The proof is complete.

\subsection{Proof of Lemma \ref{lem:generator-comm}} 

We first show the convergence part and then rule out the Zeno behaviors.  By the same procedure with the proof in Lemma \ref{lem:generator}, we can put \eqref{sys:event-generator} into the following form:
\begin{align}\label{sys:osg-reduced-event}
	\begin{split}
		\dot{\hat z}_1&=-\alpha M_1^\top {\bm \Pi}\\
		\dot{\hat z}_2&=-\alpha M_2^\top {\bm \Pi}-\beta M_L \hat z_2 + \alpha M_L \hat z_2-M_L\hat v_2+\Delta_1\\
		\dot{\hat v}_2&=-\alpha M_L {\hat v}_2+\alpha^2 M_L \hat z_2-\alpha^2 M_2^\top {\bm \Pi} + \alpha \Delta_2
	\end{split}
\end{align}
with $\hat v_1\equiv {\bf 0}$, $\Delta_1\triangleq (\beta-\alpha) M_L M_2^\top \bar z+M_LM_2^\top \bar v$, and $\Delta_2\triangleq M_LM_2^\top \bar v -\alpha M_LM_2^\top \bar z$. 

We take the time derivative of $W_0$  defined in the proof of Lemma \ref{lem:generator} along the trajectory of \eqref{sys:osg-reduced-event} and have 
\begin{align*}
	\dot{W}_0&\leq -2\eta W_0+\hat z_2^\top \Delta_1+\frac{2}{\alpha^2}\hat v_2^\top \Delta_2\\
	&\leq -\eta W_0+\frac{||\Delta_1||^2}{2\eta}+\frac{||\Delta_2||^2}{ \alpha \eta}
\end{align*}
Recalling the expressions of $\Delta_1$ and $\Delta_2$, we have 
\begin{align*}
	\dot{W}_0& \leq -\eta W_0+ \tilde k_e||\mbox{col}(\bar z,\,\bar v)||^2\\
	& \leq -\eta W_0+ N \tilde k_e(\tilde c_0+\tilde c_1e^{-\tilde \gamma t})^2
\end{align*}
holds for some constant $\tilde k_e>0$. Solving this differential inequality or recalling Lemma 9.8 in \cite{khalil2002nonlinear}, one can conclude that $W_0(t)$ is uniformly ultimately bounded with an ultimate bound $\tilde \kappa \tilde c_0^2$ for some constant $\tilde \kappa>0$. Recalling the definition of $W_0$, one can directly conclude that $z_i-y^*$ converges to a ball centered at $0$ with a radius proportional to $\tilde c_0$. Particularly, when $\tilde c_0=0$, we have 
\begin{align}\label{eq:lem-event-generator-W}
	W_0(t)&\leq W_0(0)e^{-\eta t}+  N \tilde k_e \tilde c_1^2 \int_{0}^t e^{-\eta(t-\tau)} e^{-2\tilde \gamma \tau} {\rm d}\,\tau \nonumber\\
	& = W_0(0)e^{-\eta t}+  \frac{N \tilde k_e \tilde c_1^2}{\eta- 2\tilde \gamma} [e^{-2\tilde \gamma t}-e^{-\eta t}] \nonumber\\
	&\leq k_{W} [e^{-2\tilde \gamma t}+e^{-\eta t}]
\end{align}
with $k_{W} \triangleq W_0(0) + \frac{N \tilde k_e \tilde c_1^2}{\eta- 2\tilde \gamma}  $. This implies  the exponential convergence of $z_i$ towards $y^*$ as $t$ tends to $\infty$. 

Next, we prove the Zeno-free property of the event-triggered optimal signal generator \eqref{sys:event-generator}. Suppose agent $i$ is triggered and broadcast its own information at $\tau^*\geq 0$. We want to determine the next time instant.  Let us consider the evolution of $\mbox{col}(\bar {z}_i(t),\bar{v}_i(t))$. Note that the right-hand side of \eqref{sys:osg-reduced} is  Lipschitz with respect to  $\mbox{col}(\hat {z}(t),\hat{v}_2(t))$ and the two terms $\Delta_1$ and $\Delta_2$ are Lipschitz with respect to $\mbox{col}(\bar {z},\bar{v})$. Jointly using the fact that $ ||\mbox{col}(\dot{\bar {z}}_i(t),\,\dot{\bar{v}}_i(t))||\leq ||\mbox{col}(\dot{\bar {z}}(t),\dot{\bar{v}}(t)||=||\mbox{col}(\dot{\hat z}(t),\dot{\hat{v}}_2(t)||$, we have
\begin{align*}
	||\mbox{col}(\dot{\bar {z}}_i(t),\,\dot{\bar{v}}_i(t))||&\leq \tilde h_M ||\mbox{col}(\hat {z}(t),\hat{v}_2(t))||+\tilde h_\Delta ||\mbox{col}(\bar {z},\bar{v})||\\
	&\leq  \sqrt{2\alpha^3} \tilde h_M \sqrt{W_0(t)}+\tilde h_\Delta ||\mbox{col}(\bar {z},\bar{v})||
\end{align*}
for some constants $\tilde h_M,\, \tilde h_\Delta>0$. 

We consider two different cases ($\tilde c_0\neq 0$ and $\tilde c_0=0$) to complete the proof. When $\tilde c_0\neq 0$, it holds that 
\begin{align*}
	||\mbox{col}(\dot{\bar {z}}_i(t),\,\dot{\bar{v}}_i(t))||&\leq \tilde h_M \sqrt{4\alpha^3 k_W} +\tilde h_\Delta \sqrt{N} (\tilde c_0+\tilde c_1)\triangleq \tilde k_u
\end{align*}
Hence, $||\mbox{col}({\bar {z}}_i(t),\,\bar{v}_i(t))||\leq \tilde  k_u (t-t^*)$. Then, the next event will not be triggered before $\tau^*+\tilde c_0/\tilde k_u$.  When $\tilde c_0=0$, we use the bound of $W_0(t)$ and conclude that
\begin{align*}
	||\mbox{col}(\dot{\bar {z}}_i(t),\,\dot{\bar{v}}_i(t))||&\leq \tilde{\hat c}_1(e^{-\frac{\eta t}{2}}+e^{-\tilde \gamma t }) =2\tilde{\hat c}_1 e^{-\tilde \gamma t }
\end{align*}
for positive $\tilde{\hat c}_1>0$. Consider the equation $2\tilde{\hat c}_1 s-\tilde c_1 e^{-\tilde \gamma s}=0$. Note that the left-hand side of this equation is monotone, it has a unique solution $\tilde \tau_1>$ by the intermediate value theorem.  For any $\tau^*<t<\tau^*+\tilde \tau_1$, it follows that
\begin{align*}
	||\mbox{col}({\bar {z}}_i(t),\,{\bar{v}}_i(t))||\leq 2\tilde{\hat c}_1 e^{-\tilde \gamma t^* }(t-\tau^*)<{\hat c}_1 e^{-\tilde \gamma t }
\end{align*}
That is, the next event will not be triggered before $\tau^*+\tilde \tau_1$.  In this way, we set $\tilde \tau_{\min}=\min\{\tilde c_0/\tilde k_u,\,\tilde \tau_1\}$ for both cases. The proof is thus complete.


\bibliographystyle{IEEEtran}
\bibliography{opt_event}

\end{document}